\documentstyle[12pt,aasms4]{article}

\begin{document}

\title{ON ELECTROSTATIC SCREENING OF IONS IN ASTROPHYSICAL PLASMAS}

\author{M. BR\"UGGEN and D.O. GOUGH}

\affil{Institute of Astronomy, Madingley Road, Cambridge CB3 0HA; and Department of Applied Mathematics and Theoretical Physics, Silver Street, Cambridge CB3 9EW, UK}

\authoremail{mbruggen@ast.cam.ac.uk}
\authoremail{douglas@ast.cam.ac.uk}

\vspace{.2in}

\begin{abstract}

There has been some controversy over the expression for the so-called `interaction energy' due to screening of charged particles in a plasma. Even in the relatively simple case of weak screening, first discussed in the context of astrophysical plasmas by Salpeter (1954), there is disagreement.
In particular, Shaviv and Shaviv (1996) have claimed recently that by not considering explicitly in his calculation the complete screening cloud, Salpeter obtained a result for the interaction energy between two nuclei separated by a distance $r$ which in the limit $r\rightarrow 0$ is only 2/3 the correct value. It appears that this claim has arisen from a fundamental misconception concerning the dynamics of the interaction. We rectify this misconception, and show that Salpeter's formula is indeed correct. 
\end{abstract}


\section{INTRODUCTION}

Screening of charged particles in a plasma is important in astrophysics because it influences both nuclear reaction rates and the equation of state of the plasma. In general, its effect is difficult to calculate. But in the case of classical weak screening of heavy ions, in which the Poisson-Boltzmann equation for the electrostatic potential can be linearized, the calculation should be relatively straightforward.

The problem was considered first by Salpeter (1954), who obtained an expression for the `interaction energy' $E_{\rm int}(r)$ between two screened nuclei separated by a distance $r$ in a dilute plasma. Recently, Shaviv and Shaviv (1996), in a paper motivated by the solar neutrino problem, challenged the procedure by which Salpeter had carried out the calculation, and disputed the formula he had derived for $E_{\rm int}(r)$ (cf DeWitt, Graboske \& Cooper 1973, Itoh, Totsuji, \& Ichimaru, 1977, Mitler 1977, Alastuey \& Jancovici 1987, Shoppa {\em et al.} 1993). The root of the issue concerns the fact that in evaluating $E_{\rm int}(r)$ Salpeter considered explicitly only the contribution to the screening cloud that is associated with just one of the screened ions. We explain why this procedure actually leads to the correct result in the weak-screening limit, showing how $E_{\rm int}(r)$ can be calculated when the entire screening cloud is taken explicitly into account.

The disputation appears to have arisen in part out of an erroneous view of what the interaction energy $E_{\rm int}$ should mean. In seeking a resolution it is therefore necessary to specify what $E_{\rm int}$ is to be used for.
The context in which we discuss the issue is that which motivated Salpeter, namely the Coulomb barrier penetration problem associated with thermonuclear reactions. This is also the context in which the recent disagreement has occurred. In this context, $E_{\rm int}$ is the potential energy $V$ appearing in the Schr\"odinger  equation describing the relative motion of the interacting ions (nuclei); so from now we shall call it $V$.
Misunderstanding is less likely to arise when discussing the equation of state, because in thermodynamics one is forced to differentiate various forms of energy.

It is common to discuss Coulomb barrier penetration in terms of the JWKB approximation to the solution of Schr\"odinger's equation. In that case, the effect of screening is simply to multiply the penetration probability by ${\rm exp}(-V_{\rm s}(0)/kT) $, where $V_{\rm s}(r)$ is the contribution to $V(r)$ from the screening, $k$ is Boltzmann's constant, and $T$ is temperature.
Thus it has become common simply to discuss the issue in terms of the value of $V_{0}\equiv V_{\rm s}(0)$. We emphasize, however, that making the JWKB approximation is quite separate from making the approximations to the screening energy: the relevance of our discussion is to the construction of the Schr\"odinger equation, irrespective of the approximations that are subsequently employed in solving it.

The Schr\"odinger equation for the relative motion between two particles is a quantum mechanical description of the nonrelativistic balance between kinetic energy and potential energy. Kinetic energy is lost or gained, to the benefit or at the expense of potential energy, $V$, which is manifest as a force on the particles.
One way of calculating that force is by using the principle of virtual work. It is therefore evident that the potential energy $V$ appearing in Schr\"odinger's equation is equal to the virtual work in a realizable motion, as Mitler (1977) has pointed out, and it is that quantity whose gradient is the relative force between the two screened particles. We demonstrate in this paper how the virtual work is correctly calculated.

After establishing the formulae for the charge distribution and the associated electrostatic potential in the weak-screening limit, we argue from a mechanistic viewpoint why Salpeter's calculation does yield the correct expression for the virtual work. In the subsequent section we reconcile $V$ with the total electrostatic  internal energy $U$ of the complete two-(screened)-particle system. There follows a discussion of the Shavivs' main objections to our argument.

Screening is an accumulation of electrons (and light ions) in the vicinity of heavy ions, resulting in a systematic deviation of the background charge distribution from that uniform state which would have been statistically the most likely had the heavy-ion charge distribution been strictly uniform. Consequently, there is a reduction of the entropy of the plasma, by an amount 
$S_{\rm int} = (U-V)/T$, which precisely accounts for the difference between the total electrostatic energy calculated with the entire screening charge distribution and the electrostatic interaction energy of each ion (and not its associated screening cloud) in the screened potential of the other. The self-energy of the screening cloud which is associated with the entropy reduction can be considered as heat, and is not available to work at constant temperature on the nuclei to modify their relative kinetic energy. Therefore it does not contribute to the potential that appears in the Schr\"odinger equation describing the relative motion of the heavy ions.

\section{WEAK SCREENING}

We consider, as did Salpeter (1954), the screening of ions $i$, considered as point particles with charges $Z_{i}e$, located at positions ${\bf r}={\bf r}_{i}$ in a nonrelativistic dilute neutral plasma at temperature $T$.
Here $e$ is the magnitude of the electron charge. The screening cloud is composed of electrons and also, in cases when the screened charge is a (usually rare) heavy ion, the relatively light ions of the plasma, whose mean charge number is $Z_{\rm p}$. It is assumed that the motion of the screened ions is slow compared with the motion of the screening particles. (For a discussion of `dynamic screening' of ions for which this assumption is not valid, see Carraro {\em et al.} 1988.) Then at any instant one may take the screening to be the continuous static equilibrium charge distribution in the electrostatic field of the ions. Associated with that distribution is an electrostatic potential $\phi ({\bf r})$ which satisfies the Poisson-Boltzmann equation, which may be written

\begin{equation}
 \nabla ^{2}\phi=4 \pi n_{\rm e}e[{\rm exp}(+\frac{e\phi}{kT})-{\rm exp}(-\frac{Z_{\rm p}e\phi}{kT})] - 4\pi \sum_{i}Z_{i}e \delta({\bf r}- {\bf r}_{i})
\end{equation}
in which $n_{\rm e}$ is the equilibrium electron number density, ${\bf r}_{i}$ is the position of the heavy ion $i$, and $\delta$ is the Dirac delta function. The reader is referred to Salpeter (1954) for a discussion of the conditions under which this equation is valid.

In the case of weak screening, defined by the condition $e\phi << kT$, equation (1) is linearized in $e\phi/kT$. The solution may then be written $ \phi({\bf r})=\sum_{i}\phi_{i}$, where $\phi_{i}$ is what the screening potential about the ion $i$ would have been had the ion been isolated from all other discrete heavy ions: it is the Debye-H\"uckel potential 
\begin{equation}
 \phi_{i}(r_{i};\kappa)=\frac{Z_{i}e}{r_{i}}{\rm exp}(-\kappa r_{i}),
\end{equation}
where $r_{i}=|{ \bf r}-{\bf r}_{i}|$ and $\kappa$ is the reciprocal of the Debye length $\lambda_{\rm D}$:

\begin{equation}
 \kappa (T)= \lambda _{\rm D}^{-1}=\left( \frac{4 \pi n_{\rm e}e^{2}( 1 + Z_{\rm p})}{kT} \right)^{1/2}.
\end{equation}
In cases where only a fraction $\alpha$ of the ions contribute to the screening, $Z_{\rm p}$ is replaced by $\alpha Z_{\rm p}$ in equation (3).
The associated contribution to the charge distribution is
\begin{equation}
 \rho_{i}=-\frac{Z_{i}e\kappa ^{2}}{4\pi r_{i}}{\rm exp}{(-\kappa r_{i})}+Z_{i}e\delta ({\bf r}-{\bf r}_{i}),
\end{equation}
the total charge distribution being $ \rho = \sum_{i}\rho_{i}$. Thus, when linearization is valid, the total screening is simply the superposition of the linearized screening of each ion considered separately. Indeed, it is sometimes convenient to refer to these two contributions to the total screening cloud separately. We shall call them cloud 1 and cloud 2.

\section{MECHANISTIC DISCUSSION OF THE SCREENED ION INTERACTION}

Consider the electrostatic interaction of the screened ions. There are four charge components to consider: ion 1, its associated screening cloud, ion 2, and its screening cloud. We need to evaluate the repulsive force ${\bf F}_{12}$ of ion 1 on ion 2. Thus we consider ion 2 in the total screened potential $\phi'(r_{1},r_{2};\kappa) = \phi - Z_{2}e/r_{2}$ of both ion 1 and the entire screening cloud.
The force is evidently ${\bf F}_{12}=-Z_{2}e{\bf \nabla}\phi'$ evaluated at ${\bf r}={\bf r}_{2}$. However, since the electrostatic potential due to the component of the screening cloud associated with ion 2 is symmetrical about ion 2, it does not contribute to the gradient at the location of ion 2, and the value of the electric field experienced by ion 2 is the same as the gradient of the screened potential of ion 1, which is what Salpeter calculated. Likewise, the value of the total potential gradient at ${\bf r}={\bf r}_{1}$ is the same as the value of the gradient of $\phi_{2}$. The force on ion 2 has magnitude 
$|{\bf F}_{12}|\equiv{\cal F}_{12}=-Z_{2}e\partial \phi'(r_{1},r_{2};\kappa) /\partial r_{1}=-Z_{2}e\partial \phi_{1}(r_{1};\kappa) /\partial r_{1}$ 
evaluated at 
$r_{1}=|{\bf r}_{2}-{\bf r}_{1}|\equiv r$
, the distance separating the two ions. The potential energy $V$ associated with the interaction is the virtual work:
\begin{equation}
V=Z_{2}e\int^{r}_{\infty}\frac{\partial \phi_{1}}{\partial r_{1}}{\rm d}r_{1}=Z_{2}e\phi_{1}(r;\kappa)=\frac{Z_{1}Z_{2}e^{2}}{r}e^{-\kappa r} ,
\end{equation}
in which the integration is carried out at constant temperature, and therefore at constant $\kappa$, and $e^{-\kappa r}$ stands for ${\rm exp}{(-\kappa r)}$. Of course, one could equally well have considered the electrostatic force on ion 1 in the screened potential of ion 2, obtaining the same result.
The screening contribution $V_{\rm s}$ to the potential energy of interaction, namely the difference between $V$ and the potential energy $Z_{1}Z_{2}e^{2}/r$ of the corresponding unscreened interaction, is
\begin{equation}
 V_{\rm s}(r)=\frac{Z_{1}Z_{2}e^{2}}{r}(e^{-\kappa r}-1),
\end{equation}
whose value at $r=0$ is $V_{0}=-Z_{1}Z_{2}e^{2}\kappa$. This formula is equivalent to that obtained by Salpeter (1954).

Of course there is also an attractive force between the screened ion 1 and the screening cloud about ion 2, whose magnitude is $ {\cal F}_{\rm cl}$. But since in the linearized approximation the screening cloud remains distributed symmetrically about its ion, none of that force is transmitted to ion 2.  
It is transmitted instead to the rest of the plasma. 
Therefore, in this method of calculation, the force of both ion 1 and its screening 
cloud on the screening cloud of ion 2 does not contribute to $V$.  We note that the forces exerted by the plasma on the two components of the screening cloud 
must be equal and opposite, because no net force can be exerted by the 
otherwise uniform plasma on the screened ion pair.

In reality, the different forces on the ion and its associated screening cloud polarize the system, causing the ion and its screening cloud to interact electrostatically, thereby reducing somewhat the magnitude of $V$. But that reduction is nonlinear in $e\phi/kT$, and is formally negligible in the weak-screening limit.

It is interesting to note that if it were to be assumed that the polarization is simply a small displacement of the screening cloud relative to its ion, without distortion, such that the effect is simply to transmit the force on the cloud to the ion, $V$ would become $Z_{1}Z_{2}e^{2}r^{-1}(1-\frac{1}{2}\kappa r)e^{-\kappa r} $, and $V_{0}$ would be $-\frac{3}{2}Z_{1}Z_{2}e^{2}\kappa$, which is the value preferred by Shaviv and Shaviv (1996). However, such an assumption is inconsistent: if the ions are labeled such that $Z_{1}\geq Z_{2}$ there is no possible nontrivial displacement of the undistorted screening cloud of ion 2 that can transmit to that ion a force equal to ${\cal F}_{\rm cl}$. Any consistent generalization of the weak-screening formula must take due account of  the distortion of each screening cloud by the electrostatic attraction of the neighbouring screened ion.

\section{THERMODYNAMICAL DISCUSSION OF THE SCREENED ION INTERACTION}

We now regard the screened ion pair discussed in the previous section as an integral part of the plasma. Our starting point is the electrostatic energy of their interaction, which contributes to the internal energy of the plasma. It is given by
\begin{equation}
 U = \frac{1}{2} \int \phi ({\bf r}) \rho ({\bf r}){\rm d}^{3}{\bf r},
\end{equation}
the integral being over all space. The components $\phi_{i}$ and $\rho_{i}$ of $\phi$ and $\rho$ are given by equations (2) and (4). Evaluation of the integral is straightforward, and is outlined in the Appendix. The result is
\begin{equation}
U(r,T)= U_{\rm int}(r,T) + U_{\infty}=\frac{Z_{1}Z_{2}e^{2}}{r}(1-{1\over 2}\kappa r )e^{-\kappa r} + U_{\infty},
\end{equation}
in which $r$ is again the separation of the ions, and $U_{\infty}(T)$ is independent of $r$. The contribution to $U$ from the screening is $U_{\rm s}=U-Z_{1}Z_{2}e^{2}/r.$

In order to evaluate the potential $V$, we apply the first law of thermodynamics:
\begin{equation}
{\rm d}U=T{\rm d}S-{\cal F}_{12}{\rm d}r,
\end{equation}
where $S$ is the entropy of the screened ion pair; it is given by
\begin{equation}
S=-\left (\frac{\partial F}{\partial T}\right )_{r}=\frac{U-F}{T} ,
\end{equation}
where $F=U-TS$ is the Helmholtz free energy. The second of equations (10) integrates at constant $r$ to
\begin{equation}
F(r,T)=T\int_{T}^{\infty}\frac{U(r,T')}{T'^{2}}{\rm d}T'.
\end{equation}
From equation (9) and the definition of $F$ one obtains ${\rm d}F=-S{\rm d}T-{\cal F}_{12}{\rm d}r$, from which it follows that
\begin{equation}
{\cal F}_{12}=-\left (\frac{\partial F}{\partial r}\right )_{T}.
\end{equation}
The force between the two screened particles is minus the isothermal gradient of the free energy, irrespective of the degree of screening, as DeWitt, Graboske and Cooper (1973) and Mitler (1977) have pointed out. It is not minus the isothermal gradient of the internal energy. Thus the potential $V$ in Schr\"odinger's equation is always $F_{\rm int}\equiv F(r,T)-F(\infty ,T)$, whether the screening is weak, intermediate or strong. With this identification, substituting the weak-screening expression (8) for $U$ into equation (11) yields $V=F_{\rm int} =Z_{1}Z_{2}e^{2}r^{-1}e^{-\kappa r}$, which recovers the final expression in equation (5), and hence Salpeter's expression (our equation 6) for $V_{\rm s}$.
The repulsive force on ion 2 in the electrostatic potential of both ion 1 and the (entire) screening cloud, according to equation (12), is given by
\begin{equation}
{\cal F}_{12}=\frac{Z_{1}Z_{2}e^{2}}{r^{2}} (1+\kappa r)e^{-\kappa r},
\end{equation}
and the entropy, which can be calculated from equations (10), is
\begin{equation}
S(r,T) = S_{\infty}(T)-\frac{Z_{1}Z_{2}e^{2}\kappa}{2 T}e^{-\kappa r},
\end{equation}
where $S_{\infty}$ is the limit of the entropy at temperature $T$ as the separation between the ions tends to infinity.

\section{DISCUSSION}

In the weak-screening limit, the Poisson-Boltzmann equation (1) is linearized, so that the screening cloud surrounding two stationary (or slowly moving) heavy ions can be represented by a superposition of the two spherically symmetrical screening clouds that would accumulate independently about each ion considered in isolation. The concentration of screening particles in the potential well of the two ions reduces the electrostatic energy of the system; it signifies an increase in order in the plasma, and therefore a reduction in entropy.

When considering the consequences of screening to barrier penetration, it is necessary to calculate the potential energy $V$ of the mechanical interaction between the ions, whose derivative with respect to the ion separation $r$ is the force between those ions. That force is purely electrostatic: the force on ion 2 is simply the product of the charge of ion 2 and the electric field due to both ion 1 and the total screening cloud. However, since the electric field of the component of the screening cloud associated with ion 2 vanishes at its centre of symmetry, where ion 2 is located, the value of the total electric field at this location, {\em and at this location only}, is the same as the value of the field due only to ion 1 and its associated screening cloud.  The electrostatic force between the ions can be calculated either as the isothermal derivative of the free energy, or as the product of the charge of ion 2 and that part of the screened charge of ion 1 that is contained in the sphere of radius $r$ about ion 1 (obtained by integrating $\rho$, given by equation 4, over the volume of the sphere) divided by $r^2$, or the equivalent expression obtained by interchanging the labels 1 and 2. The potential from which that force is derived is the potential $V$ that appears in Schr\"odinger's equation; it is the same as that adopted by Salpeter.

In his referee's report on the preceding sections of this paper, G. Shaviv, in consultation with N. Shaviv, has claimed that our argument is fallacious. We address here just his principal two objections, in the hope of illucidating the physics involved.

The essence of the first objection is based on the fact that, in what Shaviv calls the adiabatic picture, the total screening energy, namely the difference between the total interaction energy $U_{\rm int}$ given by equation (A3) and the interaction energy $Z_{1}Z_{2}e^{2}/r$ of the bare ions, is (at zero ion separation) 3/2 of the value of $V_{\rm s}(0)$ given by equation (6). Shaviv complains that he cannot understand why that should not imply that our value of $V_{0}=V_{\rm s}(0)$ must be multiplied by the factor 3/2. 

The reason why it must not can be understood as follows: The difference between $U_{\rm int}$ and $V$ is the interaction energy $U_{\rm cl}$ between the two cloud components, associated with which is a contribution to the entropy of $S_{\rm cl} = T^{-1}U_{\rm cl}$ of the screened-ion system. So far as the mechanics of the ions is concerned, that cloud-cloud interaction energy is essentially heat, and it is not available to contribute to the force between the ions in the isothermal motion.   The plasma exerts a force on each screening cloud which balances the electrostatic force exerted on the cloud by the other screened ion.  The energy $U_{\rm cl}$ is exchanged with the background plasma, which acts as a heat bath. It is the remaining energy, $U_{\rm int} - U_{\rm cl}$, namely the interaction free energy $F_{\rm int}$, whose isothermal derivative with respect to $r$ is the mutual force between the ions, and which must therefore be identified with the potential $V$ in Schr\"odinger's equation.

We suspect that part of the confusion experienced by Shaviv has arisen from an  incorrect assumption that the actual motion of the screened system is adiabatic. To be sure, one can calculate the force between the ions by considering a virtual adiabatic displacement. The force is evidently the derivative with respect to $r$ of the total electrostatic interaction internal energy $U_{\rm int}$ at constant entropy $S$, because in such a virtual displacement 
there is no heat exchange with the plasma. The outcome is, of course, the same as that given by equation (12), since $ (\partial U/ \partial r)_{S}=(\partial F/ \partial r)_{T}$.
This may be verified by differentiating expression (A3) for $U_{\rm int}$ at constant $S$ (which is given by equation 14), yielding equation (13). In a virtual displacement corresponding to that derivative, temperature varies, so the displacement does not correspond to the actual motion. The real motion is isothermal, because the screened heavy ions move slowly in the heat bath provided by the rapidly moving electrons and light ions of the surrounding isothermal plasma. Therefore, in computing from this result the potential $V$, the integral $\int {\cal F}_{12}dr$ must be carried out at constant temperature, which yields equation (5).

We observe that the result of Shaviv and Shaviv (1996) is obtained by replacing the force with the isothermal derivative of $U_{\rm int}$. That is evidently an incorrect procedure, because it ignores the heat exchange with the plasma. Moreover, it would predict that the force between the screened ions is attractive when $\kappa r > 1 + \sqrt{3}$. It is most straightforward to see that that cannot be so by considering the special case $Z_{1}=Z_{2}$; then the effective charges of each screened ion experienced by the other are evidently identical, and the force between them must therefore necessarily always be repulsive. 

Shaviv's second objection rests on the mistaken belief that in our calculation of the force exerted by ion 1 on ion 2 we do not take into account the electrostatic potential of the screening cloud associated with ion 2, and that consequently that force is not necessarily equal and opposite to the force exerted by ion 2 on ion 1. He points out that the force on a bare ion 2 in the electrostatic field of the screened ion 1 is different from the force on ion 1 in the electrostatic field of the bare ion 2, which, to the best of his knowledge, contradicts Newton's third law. 

We do not take issue with the second of these statements, because it does not apply to our calculation. But we do take issue with the first. That we do not ignore the contribution to the electrostatic potential from the screening cloud associated with ion 2 is most readily seen in the Appendix, for its contribution to the interaction energy is explicitly the third term of the integrand in equation (A1). We point out in section 3 why the electrostatic potential of screening cloud 2 does not contribute to the force on ion 2: we did take it into account, but it contributed nothing. And likewise, the potential of cloud 1 does not contribute to the force on ion 1. Contrary to Shaviv's claim, the situation described by us is symmetric with respect to the two ions, and action and reaction are equal and opposite. Indeed, ${\bf F}_{12}=-Z_{2}e\partial \phi_{1}(r_{1};\kappa)/\partial r_{1}|_{r_{1}=r}\partial r/\partial {\bf r}_{2}$ and ${\bf F}_{21}=-Z_{1}e\partial \phi_{2}(r_{2};\kappa)/\partial r_{2}|_{r_{2}=r}\partial r/\partial {\bf r}_{1}$, where, as previously, $r=|{\bf r}_{2}-{\bf r}_{1}|$; and since $Z_{1}e\phi_{2}(r;\kappa)=Z_{2}e\phi_{1}(r;\kappa)$ and $\partial r/\partial {\bf r}_{1}=-\partial r/\partial {\bf r}_{2}$ it follows that ${\bf F}_{21}=-{\bf F}_{12}$, in agreement with Newton's third law.  It is straightforward to demonstrate that the force of the plasma on the screening cloud of ion 2 is equal and opposite to the force of the plasma on the screening cloud of ion 1, confirming that the plasma exerts no net force on the screened ion pair.

Finally, one must bear in mind that although the principles of our discussion are valid quite generally for essentially static screening, the detailed formulae are valid only in the limit of weak screening. In reality, each ion together with its screening cloud are polarized by the presence of the other, which modifies the interaction free energy. Moreover, the polarization depends not only on the separation of the ions, but also on their velocities. To take due account of the polarization, a proper quantum mechanical description of the electron distribution is required. That is beyond the scope of the present discussion.

\section{CONCLUSION}

In the weak-screening limit, Salpeter's expression, essentially equation (5), for the interaction energy that appears in the Schr\"odinger equation governing the relative motion of two screened ions is correct. In particular, the contribution from screening at zero separation is $ V_{0}=-Z_{1}Z_{2}e^{2}/ \lambda_{\rm D} $, where $Z_{1}e$ and $Z_{2}e$ are the ionic charges and $\lambda_{\rm D}$ is the Debye length of the light component of the plasma. The more recent incorrect suggestion that Salpeter's value for $V_{0}$ should be multiplied by a factor 3/2 seems to have arisen from the false assumption that the electrostatic force on a screening cloud due to the other screened ion, evaluated in the linearized approximation, is transmitted directly to the ion it surrounds, rather than to the surrounding plasma, and from a failure to appreciate that heat is exchanged between the screening cloud and the surrounding isothermal plasma as the distance $r$ between the ions varies. When the heat exchange is correctly taken into account, it transpires that it is Helmholtz free energy, and not internal energy, that must be identified with the potential energy $V$ in Schr\"odinger's equation. Not only do the free energy and the internal energy of the screening differ in magnitude at zero ion separation -- $F_{\rm int}=\frac{2}{3} U_{\rm int}$ at $r=0$ -- but they also have different functional dependences on $r$. Therefore, to use the interaction internal energy $U_{\rm int}$ rather than the interaction free energy $F_{\rm int}$ for the potential $V$ in Schr\"odinger's equation leads to yet further errors when one includes corrections from the Liouville--Green expansion about the JWKB approximation to the barrier penetration probability.

\acknowledgements

We are grateful to D. Lynden-Bell for illuminating discussions. We thank our first referee, G. Shaviv, for insisting that our discussion in sections 1--4 does not make the case, thereby causing us to add section 5 which we trust illucidates the matter satisfactorily. We also thank H. DeWitt, the second referee, for stating unambiguously that our discussion is correct.

\appendix
\section{APPENDIX}

After making the substitutions $\phi=\phi_{1}+\phi_{2}$ and $\rho=\rho_{1}+\rho_{2}$ in equation (7), and using equations (2) and (4) for $\phi_{i}$ and $\rho_{i}$, one can identify two distinct groups of terms: those representing the sum $U_{\infty}$ of the electrostatic energies of each of the two screened ions considered in isolation, and those representing the interaction energy between the two. The former does not vary with the separation $r$ between the ions, and therefore makes no contribution to either the force between the ions or the free-energy excess $F_{\rm int}$ over the free energy at infinite ion separation. Therefore we need consider only the latter group of terms. It is given by
\begin{equation}
 U_{\rm int} =  \frac{1}{2}Z_{1}Z_{2}e^{2}\int\left \{ -\frac{2\kappa^{2}}{4\pi r_{1}r_{2}}e^{-\kappa (r_{1}+r_{2})}\nonumber \\ +\left [ \frac{e^{-\kappa r_{1}}}{r_{1}}\delta ({\bf r}-{\bf r}_{2})+\frac{e^{-\kappa r_{2}}}{r_{2}}\delta ({\bf r}-{\bf r}_{1})\right] \right \}{\rm d}^{3}{\bf r}, 
\end{equation}
where $r_{1}$ and $r_{2}$ are the distances of the volume element of integration from the respective ions, and the integration is over all space. To evaluate the first term, $U_{\rm cl}$, which represents the excess of the electrostatic self-energy of the complete screening cloud over the value when the ions are infinitely separated, it is convenient to adopt ellipsoidal coordinates $(\xi, \eta, \phi)$, where now $\phi$ is the azimuthal angle about the axis upon which the ions lie, and $ r \xi = r_{1}+ r_{2} $, $ r \eta = r_{1}- r_{2}$. The appropriate part of the right-hand side of equation (A1) then reduces to

$$U_{\rm cl}=-\frac{1}{4}Z_{1}Z_{2}e^{2}\kappa^{2}r\int^{1}_{-1}{\rm d}\eta \int^{\infty}_{1}e^{-\kappa r \xi}{\rm d}\xi = -\frac{1}{2}Z_{1}Z_{2}e^{2}\kappa e^{-\kappa r}. \eqno(A2)$$
The remaining contribution is the sum of the electrostatic energies of each of the two ions in the screened potential of the other. Its value is evidently $Z_{1}Z_{2}e^{2}r^{-1} e^{-\kappa r}$. Hence 
$$
U_{\rm int} = Z_{1}Z_{2}e^{2}r^{-1}(1-\frac{1}{2}\kappa r)e^{-\kappa r}.
\eqno(A3)$$
For completeness, we record that $U_{\infty}=-\frac{3}{4}(Z_{1}^{2}+Z_{2}^{2})e^{2}\kappa$. The total electrostatic internal energy $U$ is $U_{\rm int}+U_{\infty}$, as stated by equation (8).


\begin{references}

\reference {A} Alastuey, A., \& Jancovici, B. 1978, \apj, 226, 1034
\reference {Car} Carraro, C., Sch\"afer, A., \& Koonin, S.E. 1988, \apj, 331, 565
\reference {Gr} DeWitt, H. E., Graboske, H. C., \& Cooper, M. S. 1973, \apj, 181, 439
\reference {It} Itoh, N., Totsuji, H., \& Ichimaru, S. 1977, \apj, 218, 477
\reference {M1} Mitler, H. E. 1977, \apj, 212, 513
\reference {Salpeter} Salpeter, E. E. 1954, Australian J. Phys., 7, 373
\reference {Shav} Shaviv, N. J., \& Shaviv, G. 1996, \apj, 468, 433
\reference {Sho} Shoppa, T. D., Koonin, S. E., \& Langanke, K., \& Seki, R. 1993, Phys. Rev. C, 48, 837

\end{references}
\end{document}